\documentclass[epj]{webofc}
\usepackage[utf8]{inputenc}
\usepackage[varg]{txfonts}   
\usepackage{booktabs}
\usepackage{afterpage}
\usepackage{xcolor}
\definecolor{darkred}{rgb}{0.4,0.0,0.0}
\definecolor{darkgreen}{rgb}{0.0,0.4,0.0}
\definecolor{darkblue}{rgb}{0.0,0.0,0.4}
\usepackage[bookmarks,linktocpage,colorlinks,
    linkcolor = darkred,
    urlcolor  = darkblue,
    citecolor = darkgreen]{hyperref}
\usepackage{subfigure}
\wocname{EPJ Web of Conferences}
\woctitle{Lattice2017}
\newcommand{\ahvp}{a_\mu^{\rm hvp}}
\newcommand{\ahlbl}{a_\mu^{\rm hlbl}}
\newcommand{\Nf}{N_{\rm f}}
\newcommand{\zv}{Z_{\rm V}}
\newcommand{\xcut}{x_0^{\rm cut}}
\newcommand{\be}{\begin{equation}}
\newcommand{\ee}{\end{equation}}
\newcommand{\bea}{\begin{eqnarray}}
\newcommand{\eea}{\end{eqnarray}}
\newcommand{\eq}[1]{eq.\,(\ref{#1})}

\begin{document}
%
\selectlanguage{english}
\title{A lattice calculation of the hadronic vacuum polarization
  contribution to $(g-2)_\mu$}
\author{
\firstname{M.} \lastname{Della Morte}\inst{1} \and
\firstname{A.} \lastname{Francis}\inst{2} \and
\firstname{A.}  \lastname{G\'erardin}\inst{3} \and
\firstname{V.}  \lastname{G\"ulpers}\inst{4} \and
\firstname{G.}  \lastname{Herdo{\'\i}za}\inst{5} \and
\firstname{G.}  \lastname{von Hippel}\inst{3} \and
\firstname{H.}  \lastname{Horch}\inst{3} \and
\firstname{B.}  \lastname{J\"ager}\inst{1,6} \and
\firstname{H.B.}  \lastname{Meyer}\inst{3,7} \and
\firstname{A.}  \lastname{Nyffeler}\inst{3} \and
\firstname{H.}  \lastname{Wittig}\inst{3,7}\fnsep\thanks{Speaker, \email{hartmut.wittig@uni-mainz.de}}
}
\institute{%
CP3-Origins, University of Southern Denmark, Campusvej 55,
5230 Odense M, Denmark
\and
Department of Physics and Astronomy, York University,
Toronto, ON, Canada, M3J1P3
\and
PRISMA Cluster of Excellence and Institut f\"ur Kernphysik,
University of Mainz, D-55099 Mainz, Germany 
\and
School of Physics and Astronomy, University of Southampton,
Southampton SO17 1BJ, UK
\and
Inst. de F{\'\i}sica Te\'orica UAM/CSIC and Dpto de F{\'\i}sica
Te\'orica, U. Aut\'onoma de Madrid, E-28049 Madrid
\and
ETH Z\"urich, Institute for Theoretical Physics,
Wolfgang-Pauli-Str. 27, 8093 Z\"urich, Switzerland 
\and
Helmholtz Institute Mainz, University of Mainz, D-55099 Mainz, Germany }
\abstract{We present results of calculations of the hadronic vacuum
  polarisation contribution to the muon anomalous magnetic
  moment. Specifically, we focus on controlling the infrared regime of
  the vacuum polarisation function. Our results are corrected for
  finite-size effects by combining the Gounaris-Sakurai
  parameterisation of the timelike pion form factor with the L\"uscher
  formalism. The impact of quark-disconnected diagrams and the
  precision of the scale determination is discussed and included in
  our final result in two-flavour QCD, which carries an overall
  uncertainty of 6\%. We present preliminary results computed on
  ensembles with $\Nf=2+1$ dynamical flavours and discuss how the
  long-distance contribution can be accurately constrained by a
  dedicated spectrum calculation in the iso-vector channel.\vspace{-0.5cm}}
\maketitle
\section{Introduction}\label{s1:intro}

The persistent deviation of $3.5-4$ standard deviations between the
direct measurement of the muon's anomalous magnetic moment
$a_\mu=\frac{1}{2}(g-2)_\mu$ and the value predicted by theory
\cite{Olive:2016xmw} may signal a limitation in the ability of the
Standard Model (SM) to provide an accurate and precise description of
particle properties. Two new experiments, E989 at Fermilab and E34 at
J-PARC, will increase the precision of the direct experimental
determination of $a_\mu$ by up to a factor of four, which calls for a
similar reduction in the uncertainty of the SM prediction. Since the
latter is dominated by the contributions from low-energy QCD, it is
particularly timely to revisit the determinations of the hadronic
vacuum polarisation and hadronic light-by-light scattering
contributions, $\ahvp$ and $\ahlbl$, respectively. Since the current
estimates for these quantities either rely on experimental data or on
model assumptions, it is highly desirable to compute them via a
first-principles approach such as lattice QCD. However, in order to be
competitive with the dispersive approach, any lattice QCD calculation
of $\ahvp$ must be able to control all sources of error at the level
of 0.5\% or better, if such studies are to have an impact on testing
the limits of the SM. By contrast, a determination of $\ahlbl$ with an
overall error of about 15\% will constitute a major step forward.

Here we report on our ongoing effort in determining $\ahvp$ in lattice
QCD with fully controlled errors. In particular, we describe our
calculation in QCD with two flavours, which is focussed on
methodology, and which has been published
in~\cite{DellaMorte:2017dyu}. We also discuss preliminary results
obtained on gauge ensembles with $2+1$ dynamical quark flavours.

\section{Methodology}\label{s2:method}

The leading hadronic vacuum polarisation (HVP) is accessible in
lattice QCD via two types of integral representations. The first is
formulated in terms of a convolution integral over Euclidean momenta
\cite{Lautrup:1971jf,Blum:2002ii} and reads
\be
  \ahvp = \left(\frac{\alpha}{\pi}\right)^2 \int_0^\infty 
  dQ^2\,K(Q^2)\hat\Pi(Q^2),
\ee
where $K(Q^2)$ is a known kernel function, and
$\hat\Pi(Q^2)=4\pi^2\left(\Pi(Q^2)-\Pi(0)\right)$ denotes the
subtracted vacuum polarisation which is obtained from the correlator
of the electromagnetic current $J_\mu=\frac{2}{3}\bar{u}\gamma_\mu u
-\frac{1}{3}\bar{d}\gamma_\mu d -\frac{1}{3}\bar{s}\gamma_\mu
s+\cdots$ via
\be\label{eq:Pimunudef}
  \Pi_{\mu\nu}(Q)=\int d^4x\,\,{\rm e}^{iQ\cdot x}\,\left\langle
  J_\mu(x)J_\nu(0) \right\rangle\equiv (Q_\mu
  Q_\nu-\delta_{\mu\nu}Q^2)\Pi(Q^2).
\ee
Alternatively, one can express $\ahvp$ via the ``time-momentum
representation'' (TMR)\,\cite{Bernecker:2011gh}, by integrating the
spatially summed correlator $G(x_0)$ multiplied by a kernel
$\tilde{K}(x_0)$ over Euclidean time according to
\bea\label{eq:TMRdef}
   & & a_\mu^{\rm hvp} = \left(\frac{\alpha}{\pi}\right)^2
   \int_0^\infty dx_0\,\tilde{K}(x_0)\,G(x_0),\quad
   G(x_0)=-a^3\sum_{\vec{x}} 
   \left\langle J_k(x)J_k(0)\right\rangle, \\
   & & \tilde{K}(x_0)=4\pi^2\int_0^\infty dQ^2\,K(Q^2)
  \left[x_0^2-\frac{4}{Q^2}\sin^2\left(\textstyle\frac{1}{2}Qx_0
  \right)\right].
\eea
A major difficulty, which is encountered in both approaches, is
associated with controlling the infrared regime of the vacuum
polarisation. In the first approach, one must determine the additive
renormalisation $\Pi(0)$ and a sufficiently precise representation of
$\Pi(Q^2)$ for values of $Q^2$ that do not exceed the muon mass by
much. While the use of partially twisted boundary conditions is
helpful\,\cite{DellaMorte:2011aa} in this regard, it does not remove
the need for simulating large lattice volumes in order to reliably
constrain $\Pi(Q^2)$ in the low-momentum
regime\,\cite{Golterman:2014ksa}.

If one employs the TMR, one is confronted with the problem of
controlling the long-distance regime of $G(x_0)$, which is difficult
owing to its rapidly rising noise-to-signal ratio. Therefore one has
to resort to some kind of model for the large-$x_0$
behaviour. Furthermore, in the light quark sector, $G(x_0)$ is
dominated by a two-pion state in a $p$-wave as
$x_0\to\infty$. Possible choices for the description of the iso-vector
component of $G(x_0)$ at large distances include a naive single
exponential function that falls off with the mass of the
$\rho$-resonance or an {\it ansatz} that supplements the single
exponential with the contribution from the two-pion state. A third
possibility is based on the Gounaris-Sakurai parameterisation
\cite{Gounaris:1968mw} of the timelike pion form factor that enters
the iso-vector spectral
function\,\cite{Meyer:2011um,Francis:2013qna}. In
ref.\,\cite{DellaMorte:2017dyu} we presented a detailed discussion of
the different strategies to control the low-energy regime.

\section{Lattice calculation, systematic effects and results}\label{s3:calcsyst}

Our results for the hadronic vacuum polarisation have been obtained
using gauge ensembles generated as part of the CLS effort. An estimate
for $\ahvp$ in two-flavour QCD, including a detailed error budget has
been published in \cite{DellaMorte:2017dyu}. This study was performed
using two degenerate flavours of O($a$) improved Wilson quarks and the
Wilson plaquette action, using the parameterisation of the clover
coefficient $c_{\rm sw}$ from \cite{impr:csw_nf2}. Extrapolations to
the physical point have been performed using data at three different
values of the lattice spacing (i.e. $a=0.049, 0.066$ and $0.079$\,fm)
and pion masses ranging from $185-495$\,MeV, always keeping $m_\pi
L\geq4$. In addition, we present preliminary results for QCD with
$\Nf=2+1$ dynamical flavours \cite{Bruno:2014jqa}, based on the
tree-level Symanzik improved gauge action with $c_{\rm sw}$ tuned
according to ref.\,\cite{Bulava:2013cta}. The current range of lattice
spacings and pion masses comprises $a=0.050, 0.065, 0.085$\,fm and
$m_\pi=200-420$\,MeV, which is similar to the two-flavour case. We
postpone the inclusion of isospin-breaking effects to future work
(see\,\cite{thiscontrib294,thiscontrib76} for pilot studies).

\begin{figure}[thb]
  \centering
  \includegraphics[width=0.66\textwidth,clip]{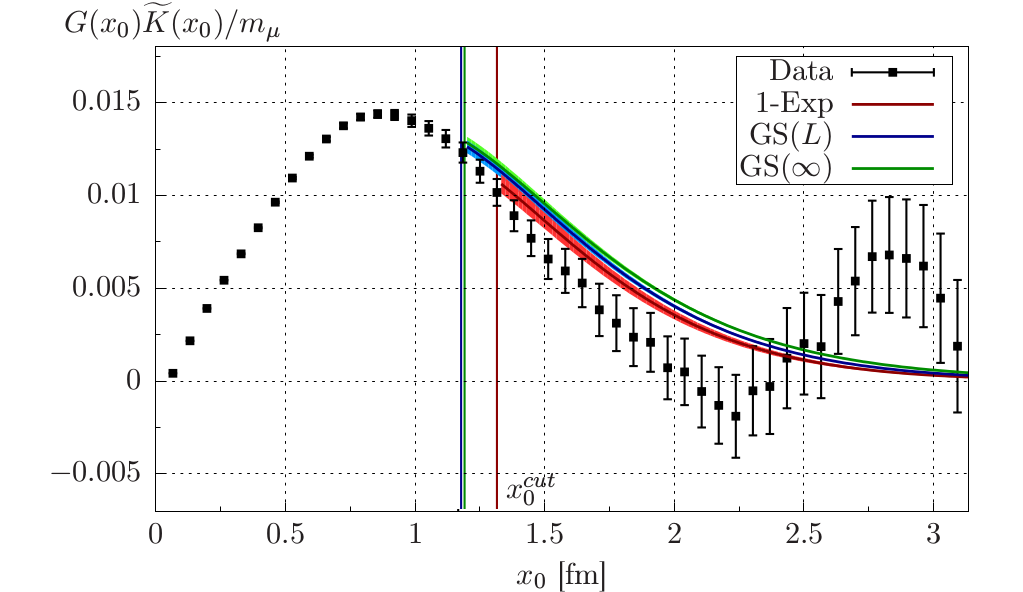}
  \vspace{-0.3cm}
  \caption{Light quark contribution to the integrand,
    $\tilde{K}(x_0)G^{ud}(x_0)$, in units of the muon mass, computed
    in two-flavour QCD at a pion mass of 185\,MeV
    \cite{DellaMorte:2017dyu}. Data are shown by black filled squares,
    while extensions of the vector correlator for $x_0 > 1.2$\,fm
    based on a single exponential and the GS parameterisation (with
    and without finite-volume correction) are represented by coloured
    bands.}
  \vspace{-0.5cm}
  \label{fig:integrand}
\end{figure}

In the following we focus on the discussion of systematic effects
arising from the long-distance behaviour of the vector correlator,
which is also connected with the issue of finite-volume effects. The
accuracy of the scale determination is also an important factor for
the overall precision of the lattice prediction of $\ahvp$. For this
purpose we restrict the discussion to the TMR. It is then convenient
to define the quark-connected contribution of flavour $f=ud, s, c$ to
the vector correlator via
\be
   G^f(x_0)= -\frac{a^3}{3}\sum_k\sum_{\vec{x}}\,q_f^2\,\zv
   \left\langle V_{k,f}^{\rm con}(x_0,\vec{x}) V_{k,f}^{\rm loc}(0)
   \right\rangle,
\ee
where $q_f$ is the quark electric charge\footnote{For $f=ud$, the
  combined iso-symmetric light quark contribution yields
  $q_{ud}^2=5/9$.}, $\zv$ denotes the renormalisation factor of the
local vector current $V_{\mu,f}^{\rm loc}(x)=
\bar\psi_f(x)\gamma_\mu\psi_f(x)$, and $V_{\mu,f}^{\rm con}(x)$
represents the conserved (point-split) current. The corresponding
contribution to the hadronic vacuum polarisation of flavour $f$ is
then given by
\be\label{eq:amufdef}
   (\ahvp)^f=\left(\frac{\alpha}{\pi}\right)^2\,\int_0^{\infty}
   dx_0\,\tilde{K}(x_0)\,G^f(x_0),\quad f=ud, s, c. 
\ee
The statistical precision of the light quark contribution
$(\ahvp)^{ud}$, which dominates the total HVP, is limited by the
exponentially growing noise-to-signal ratio of the integrand in
\eq{eq:amufdef}. This is shown in Figure~\ref{fig:integrand} where we
have plotted the product $\tilde{K}(x_0)\,G^{ud}(x_0)$ in units of
$m_\mu$ versus the Euclidean time separation $x_0$\,[fm]. Above a
certain value, i.e. $\xcut\approx 1.3$\,fm, the signal has
deteriorated to such an extent that one has to resort to a model for
$G^{ud}(x_0)$. The various coloured bands compare the extension based
on the single exponential and the Gounaris-Sakurai (GS)
parameterisation. Overall, one finds that both types of extension
yield consistent results for $(\ahvp)^{ud}$ within errors.

It is also possible to determine finite-volume corrections based on
the GS model.
%
%
To be more explicit, it is useful to start with the isospin
decomposition of the vector correlator, i.e.
\be\label{eq:isospin}
   G(x_0)=G^{\rho\rho}(x_0)+G^{I=0}(x_0),\quad
   G^{\rho\rho}(x_0)={\textstyle\frac{9}{10}} G^{ud}(x_0),
\ee
where $G^{\rho\rho}(x_0)$ denotes the iso-vector contribution, which
is proportional to the connected light-quark contribution
$G^{ud}(x_0)$. In infinite volume the iso-vector part is given by
\be\label{eq:Grhorhoinfty}
   G^{\rho\rho}(x_0,\infty)=\int_{2m_\pi}^{\infty}d\omega\,\omega^2
   \rho(\omega^2)\, e^{-\omega|x_0|}
   =\frac{1}{48\pi^2}\int_{2m_\pi}^\infty\,d\omega\,\omega^2
   \left(1-\frac{4m_\pi^2}{\omega^2}\right)^{3/2}\, 
   |F_\pi(\omega)|^2\,e^{-\omega|x_0|},
\ee
where $\rho(\omega^2)$ is the (continuous) spectral function, and
$F_\pi(\omega)$ denotes the pion form factor in the timelike
regime. In a finite volume, characterised by a box size $L$, one
encounters a discrete spectrum, i.e.
\be\label{eq:GrhorhoL}
   G^{\rho\rho}(x_0,L)\stackrel{x_0\to\infty}{=} \sum_n\,|A_n|^2
   e^{-\omega_n x_0},\quad \omega_n=2\sqrt{m_\pi^2+k^2}.
\ee
The energies $\omega_n$ are related to the scattering momentum via the
L\"uscher condition \cite{Luscher:9091}:
\be
 \delta_{1}(k)+\phi(q)= 0\,\hbox{mod}\,\pi,\quad q=\frac{kL}{2\pi}.
\ee
In the inelastic region the amplitudes $A_n$ can be expressed in
terms of the timelike pion form factor \cite{Meyer:2011um}
\be\label{eq:AnFpi}
   |A_n|^2=\,\frac{2k^2}{3\pi\omega_n^2}\,
   \frac{|F_\pi(\omega_n)|^2}{k\phi^\prime(k)+k\delta_1^\prime(k)}.
\ee
Given input data for $F_\pi(\omega)\equiv|F_\pi(\omega)|
e^{i\delta_1(k)}$, one may then determine the finite-volume shift by
forming the difference $G^{\rho\rho}(x_0,\infty)-G^{\rho\rho}(x_0,L)$
and inserting it into \eq{eq:TMRdef}. In the absence of any direct
calculation of $\omega_n$ and $|A_n|$, one can resort to the GS
parameterisation of $F_\pi(\omega)$ in terms of the resonance mass
$m_\rho$ and the width $\Gamma_\rho$. Both parameters can be
determined from lattice data via the following procedure: In a first
step, the mass of the ground state is identified with the $\rho$-meson
mass extracted from a smeared vector correlation function computed in
an auxiliary lattice calculation via a two-state fit with the lowest
level set to $2\sqrt{m_\pi^2+(2\pi/L)^2}$. Then, by inserting the GS
model for $|F_\pi|$ and $\delta_1$ into \eq{eq:AnFpi} one obtains an
expression for the correlator $G^{\rho\rho}(x_0)$ in terms of the GS
parameters $(m_\rho, \Gamma_\rho)$. Fitting $G^{\rho\rho}(x_0)$ to the
form in \eq{eq:GrhorhoL} with $m_\rho$ fixed to the estimate extracted
from the smeared correlator yields the width parameter
$\Gamma_\rho$. During this fit one inserts the GS parameterisation for
$\delta_1$ into the L\"uscher condition and determines the scattering
momenta $k$ of excited states in the iso-vector channel, which, in
turn, yield the corresponding energy levels and matrix elements
$\omega_n$ and $|A_n|^2$. In this way one can evaluate
$G^{\rho\rho}(x_0,L)$ in \eq{eq:GrhorhoL} for, say, a handful of
states. The procedure is completed by inserting the GS
parameterisation into \eq{eq:Grhorhoinfty}, which yields the
correlator $G^{\rho\rho}(x_0,\infty)$. Since the iso-scalar part in
\eq{eq:isospin} is sub-dominant, one may approximate it by a single
exponential whose fall-off is given by $m_\omega\approx m_\rho$. We
have determined the finite-volume shift in $(\ahvp)^{ud}$ according to
the above procedure, for all of our two-flavour ensembles with pion
masses $m_\pi\leq 340$\,MeV. Unsurprisingly, the biggest correction of
$+3$\,\% was encountered at our smallest pion mass of 185\,MeV.

\begin{figure}[thb]
  \centering
  \includegraphics[width=0.9\textwidth,clip]{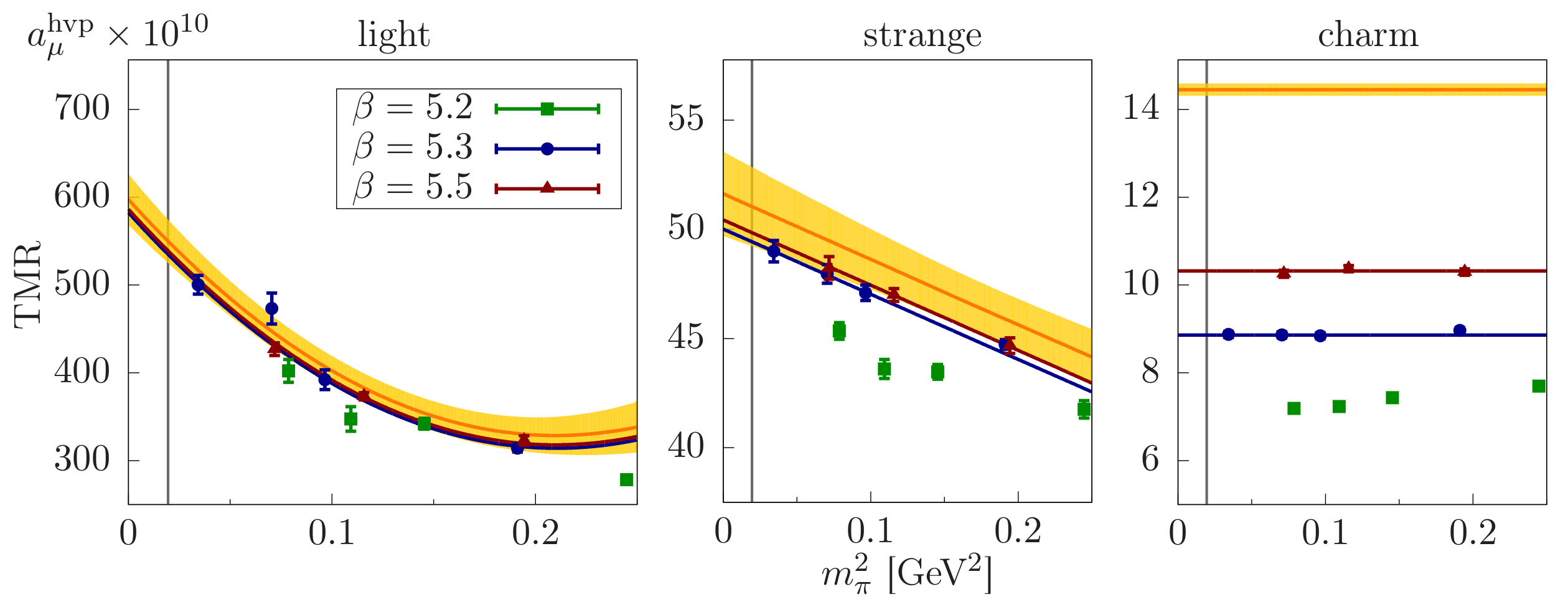}
  \vspace{-0.3cm}
  \caption{Combined chiral and continuum extrapolations of the
    individual quark flavour contributions to $\ahvp$ computed using
    the TMR in two-flavour QCD \cite{DellaMorte:2017dyu}. The yellow
    band represents the pion mass dependence in the continuum
    limit. The vertical line indicates the physical value of the pion
    mass.}
  \label{fig:extrap}
\vspace{-0.5cm}
\end{figure}

Another source of uncertainty that has received relatively little
attention in most of the earlier calculations of $\ahvp$ arises from
the uncertainty in the lattice scale. Although $\ahvp$ is
dimensionless, there are two ways in which its determination in
lattice QCD introduces a scale dependence. Firstly, the muon mass
$m_\mu$ enters the kernel function $\tilde{K}(x_0)$ via the
dimensionless combination $x_0 m_\mu$. Secondly, the masses of the
dynamical quarks enter implicitly via the lattice evaluation of the
vector correlator. Therefore, $\ahvp$ can be thought of as a function
in the dimensionless combinations $M_\mu\equiv m_\mu/\Lambda,
M_u\equiv m_u/\Lambda, M_d\equiv m_d/\Lambda,\ldots$, where $\Lambda$
is the quantity that sets the lattice scale. The scale setting error
$\Delta\Lambda$ then induces a corresponding uncertainty in $\ahvp$,
i.e.
\be
   \Delta\ahvp = \bigg|\Lambda\frac{d\ahvp}{d\Lambda}\bigg|\,
   \frac{\Delta\Lambda}{\Lambda} = \bigg|
   M_\mu\frac{\partial\ahvp}{\partial M_\mu} + \sum_{f=1}^{N_{\rm
       f}}M_{\rm f}\frac{\partial \ahvp}{\partial M_{\rm f}} \bigg|\,
   \frac{\Delta\Lambda}{\Lambda}.
\ee
It is useful to replace $M_u, M_d\,\ldots$ and their derivatives by
suitable meson masses in units of $\Lambda$. In the isospin limit the
above expression can then be rewritten as
\be
   {\Delta \ahvp} = \bigg|M_\mu\frac{\partial \ahvp}{\partial M_\mu} +
   M_\pi\frac{\partial \ahvp}{\partial M_\pi}+ M_{\rm K}\frac{\partial
     \ahvp}{\partial M_{\rm K}}+\ldots \bigg|\,
   \frac{\Delta\Lambda}{\Lambda}.
\ee
When working with the TMR one can determine the derivative term
involving the muon mass via \cite{DellaMorte:2017dyu}
\be
   M_\mu\frac{\partial a_\mu^{\rm hvp}}{\partial M_\mu} = -\ahvp
   +\left(\frac{\alpha}{\pi}\right)^2\int_0^\infty 
   dx_0\,G(x_0)\,J(x_0),\quad
   J(x_0)=x_0\tilde{K}^\prime(x_0)-\tilde{K}(x_0),
\ee
where the kernel $J(x_0)$ can be easily computed using the series
expansion of $\tilde{K}(x_0)$ from appendix~B in
\cite{DellaMorte:2017dyu}. The derivative w.r.t. the pion mass $M_\pi$
can be estimated from the slope of the chiral extrapolation of
$(\ahvp)^{ud}$ at $m_\pi=m_\pi^{\rm phys}$ (see left panel of
Figure\,\ref{fig:extrap}). With these results at hand one finds
\be
   \frac{\Delta \ahvp}{\ahvp} = \Bigg|
   \underbrace{\frac{M_\mu}{\ahvp}\frac{\partial \ahvp}{\partial M_\mu}}_{\displaystyle{1.8}}+
   \underbrace{\frac{M_\pi}{\ahvp}\frac{\partial \ahvp}{\partial M_\pi}}_{\displaystyle{-0.18(6)}}
   \Bigg|\,\frac{\Delta\Lambda}{\Lambda}.
\ee
The factor multiplying the scale setting uncertainty is thus dominated
by the contribution from the muon mass, with only a 10\% reduction
coming from the light quarks. Heavier quark flavours are likely to
have an even smaller effect. The lesson one can draw from this
analysis is the fact that the proportionality between the relative
uncertainties of $\ahvp$ and the lattice scale $\Lambda$ is a number
of order~1. Therefore, the lattice scale must be known to within a
fraction of a percent, if one is to reach the precision goal in the
determination of $\ahvp$.

We have subjected our data for the various flavour contributions to
$\ahvp$ to combined chiral and continuum extrapolations, using a
variety of different {\it ans\"atze}, such as
\be
   \begin{array}{ll}
   \hbox{Fit~A:}\quad{\alpha_1}+{\alpha_2} m_\pi^2
                    +{\alpha_3} m_\pi^2\ln m_\pi^2 +{\alpha_4}a,\quad & 
   \hbox{Fit~B:}\quad{\beta_1}+{\beta_2} m_\pi^2
                    +{\beta_3} m_\pi^4 +{\beta_4} a, \\
   \hbox{Fit~C:}\quad{\gamma_1}+{\gamma_2} m_\pi^2
                    +{\gamma_3} a, \quad &
   \hbox{Fit~D:}\quad{\delta_1}+ {\delta_2} a.
   \end{array}
\ee
Since we did not include the O($a$) improvement term in the vector
currents, we expect that the leading cutoff effects are linear in the
lattice spacing, which accounts for the term of order~$a$ in the above
expressions. Fits~A and~B each contain a term allowing for a curvature
in the chiral behaviour of $\ahvp$. Other models describing the pion
mass dependence contain terms that diverge in the chiral limit, such
as $1/m_\pi^2$ or $\ln m_\pi^2$. While the latter is only justified in
the region where $m_\pi<m_\mu$, inverse powers of $m_\pi^2$ may
over-amplify the pion mass dependence around the physical pion mass,
as was noted in \cite{Golterman:2017njs}. We have therefore excluded
singular {\it ans\"atze} in our final analysis. After adding the
contributions from the light, strange and charm quarks, we arrive at
our final estimate for $\ahvp$ \cite{DellaMorte:2017dyu}:
\be\label{eq:final}
  \ahvp = (654\pm32_{\rm\,stat}\pm17_{\rm\,syst}\pm10_{\rm\,scale}
  \pm7_{\rm\,FV}\,{}^{+\phantom{1}0}_{-10}{}_{\rm\,disc})\cdot10^{-10},
\ee
where the statistical and systematic errors have been estimated via
the ``extended frequentist's method'' which combines the standard
bootstrap technique with a number of procedural variations in the
analysis. The remaining systematic errors refer to the scale
uncertainty, the error assigned to the estimation of the finite-volume
shift and the upper bound on the effect of including
quark-disconnected diagrams. The calculation of the latter on a subset
of gauge ensembles is described in appendix~D of
ref.~\cite{DellaMorte:2017dyu}. A crucial ingredient for obtaining
accurate results is the cancellation of stochastic noise between the
contributions of the light and strange quark flavours, discussed
in\,\cite{Francis:2014hoa}.

\begin{figure}[t]
  \centering
  \includegraphics[width=0.5\textwidth,clip]{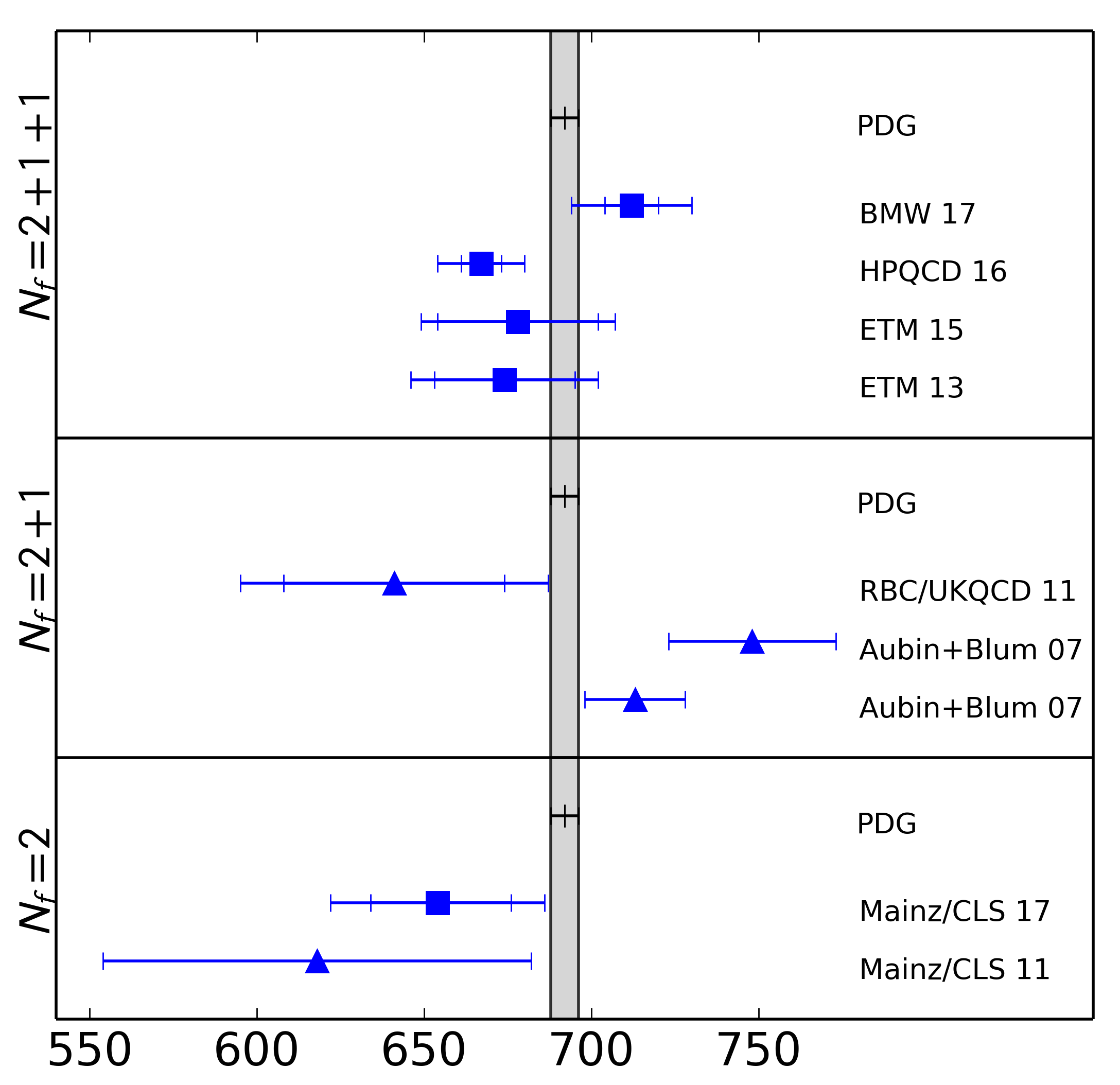}
  \caption{\label{fig:compare} Compilation of results for the hadronic vacuum polarisation
    contribution $\ahvp$ in units of $10^{-10}$. The three panels
    represent calculations with different numbers of quarks in the
    sea. Squares denote estimates including the contributions from $u,
    d, s, c$ quarks in the valence sector, while triangles represent
    results for $u, d, s$ quarks only. The meaning of the labels is:
    Mainz/CLS\,11~\cite{DellaMorte:2011aa},
    Mainz/CLS\,17~\cite{DellaMorte:2017dyu}, Aubin+Blum\,07
    \cite{Aubin:2006xv}, RBC/UKQCD\,11 \cite{Boyle:2011hu}, ETM\,13
    \cite{Burger:2013jya}, ETM\,15 \cite{Burger:2015hdi}, HPQCD\,16
    \cite{Chakraborty:2016mwy} and BMW\,17 \cite{thiscontrib221}. See
    \cite{thiscontrib205} for another recent calculation. The
    vertical band represents the result from dispersion theory.}
  \centering
  \includegraphics[width=0.7\textwidth,clip]{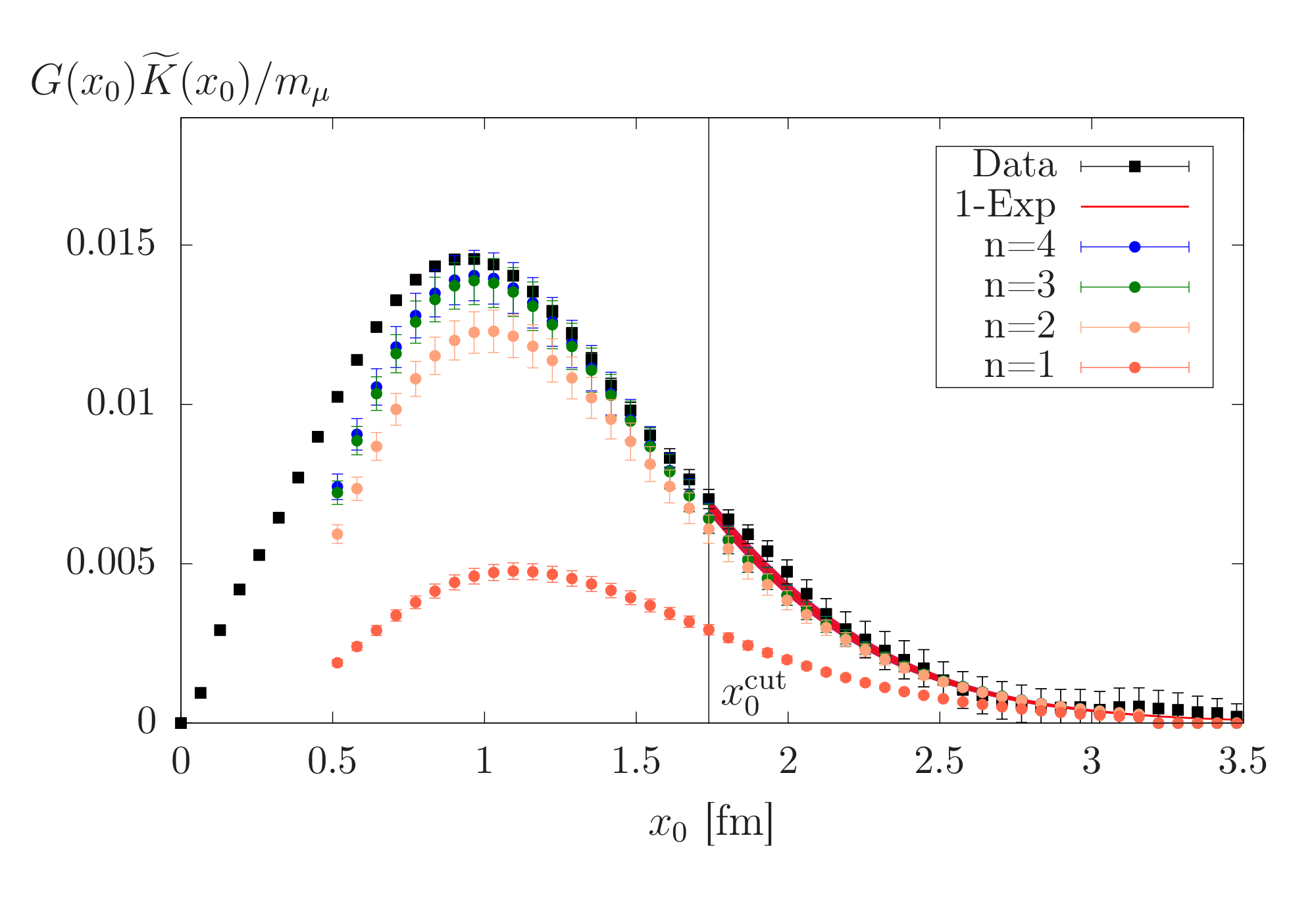}
  \vspace{-0.5cm}
  \caption{\label{fig:integrand2p1} Preliminary results for the light
    quark contribution to the integrand, $\tilde{K}(x_0)G^{ud}(x_0)$,
    in units of $m_\mu$, computed for $\Nf=2+1$ at $m_\pi=200$\,MeV.
    Data points are shown by black filled squares. The red circles
    denote the two-pion contribution to the iso-vector correlator
    $G^{\rho\rho}$, with the remaining coloured points showing the
    accumulated contributions from the higher excited states.}
\vspace{-1.0cm}
\end{figure}

A compilation of results for $\ahvp$ from different collaborations is
shown in Figure\,\ref{fig:compare}. Overall there is good agreement
among different groups, with the largest observed deviations between
individual calculations amounting to about 2.5 standard deviations. It
is clear, though, that the overall errors have to be further reduced
in order to be competitive with the dispersive analysis.

\afterpage{\clearpage}


We have started to compute $\ahvp$ on a set of CLS ensembles generated
with $\Nf=2+1$ flavours of dynamical quarks. In order to combat the
problem of topology freezing, a large fraction of the ensembles have
open boundary conditions \cite{Schaefer:2010hu,Luscher:2011kk}, which
precludes the determination of the vacuum polarisation via the
four-dimensional Fourier transform of \eq{eq:Pimunudef}. Since the TMR
is independent of the type of boundary conditions, we exclusively
focus on this approach in all our calculations for $\Nf=2+1$. One
important additional ingredient relative to the earlier two-flavour
calculation is the use of the O($a$) improved versions of the local
and point-split vector currents \cite{Harris:2015vfa}.

In Figure\,\ref{fig:integrand2p1} we show once more the integrand of
\eq{eq:TMRdef}, computed at a pion mass of 200 MeV at
$a=0.065$\,fm. The data confirm that a single exponential provides a
good approximation of the long-distance behaviour of the integrand.
We also display the results of a state-of-the-art calculation aimed at
determining the energies $\omega_n$ and corresponding matrix elements
$A_n$ of the vector current for the $n^{\rm th}$ energy eigenstate in
the iso-vector channel. This not only allows for a precise direct
calculation of the long-distance behaviour of $G^{\rho\rho}(x_0,L)$
but also provides the determination of the timelike pion form factor
that is necessary to compute the finite-volume shift. An account of
our method has been provided in \cite{thiscontrib316}. The results
from the spectrum calculation demonstrates that the iso-vector
correlator $G^{\rho\rho}(x_0)$ is saturated by the first four states
in that channel, for $x_0\gtrsim1.7$\,fm. It is also interesting to
note that the two-pion state starts to dominate the correlator for
$x_0\gtrsim3$\,fm. The explicit spectrum calculation provides a highly
accurate and precise determination of the vector correlator at long
distances, especially since the error grows only linearly with $x_0$,
in contrast to the exponential error growth of $G^{\rho\rho}$ itself.

\smallskip
\par\noindent{\bf Acknowledgments:} We thank Ben H\"orz for providing
the data of the spectrum calculation shown in
Figure\,\ref{fig:integrand2p1}. Our calculations were performed on the
HPC Clusters Wilson, Clover and MOGON-II at the University of Mainz,
as well as on the JUQUEEN computer at NIC, J\"ulich (project HMZ21).

\bibliography{lattice2017}

\end{document}